\documentstyle[times,emulateapj]{article}
\begin{document}

\title{THE FORMATION OF BROWN DWARFS AS EJECTED STELLAR EMBRYOS}


\centerline{
Bo Reipurth$^1$ and Cathie Clarke$^2$
}
 
\bigskip
\centerline{1: Center for Astrophysics and Space Astronomy,}
\centerline{University of Colorado, Boulder, CO 80309}
\centerline{(reipurth@casa.colorado.edu)  }

\bigskip
\centerline{2: Institute of Astronomy,}
\centerline{Madingley Road, Cambridge CB3 0HA, UK}
\centerline{(cclarke@ast.cam.ac.uk) }



\centerline{Accepted by the Astronomical Journal}

\begin{abstract}

We conjecture that brown dwarfs are substellar objects because they
have been ejected from small newborn multiple systems which have
decayed in dynamical interactions. In this view, brown dwarfs are
stellar embryos for which the star formation process was aborted
before the hydrostatic cores could build up enough mass to eventually
start hydrogen burning.  The disintegration of a small multiple system
is a stochastic process, which can be described only in terms of the
half-life of the decay. A stellar embryo competes with its siblings in
order to accrete infalling matter, and the one that grows slowest is
most likely to be ejected. With better luck, a brown dwarf would
therefore have become a normal star. This interpretation of brown
dwarfs readily explains the rarity of brown dwarfs as companions to
normal stars (aka the ``brown dwarf desert''), the absence of wide
brown dwarf binaries, and the flattening of the low mass end of the
initial mass function. Possible observational tests of this scenario
include statistics of brown dwarfs near Class~0 sources, and the
kinematics of brown dwarfs in star forming regions while they still
retain a kinematic signature of their expulsion.  Because the ejection
process limits the amount of gas brought along in a disk, it is
predicted that substellar equivalents to the classical T Tauri stars
should be very rare.

\end{abstract}  

\keywords{stars: low-mass, brown dwarfs -- stars: formation -- stars:
pre-main sequence -- stars: luminosity function, mass function --
binaries: general -- instabilities}

\section{INTRODUCTION}

The past few years have seen a wealth of discoveries of brown dwarfs,
whose bona fide nature as substellar objects have been confirmed by
the presence of lithium in their spectra (Rebolo, Mart\'\i n,
Magazz\`u 1992; Mart\'\i n, Basri, Zapatero Osorio 1999). Numerous
brown dwarfs have been found in clusters like the Pleiades
(e.g. Stauffer et al. 1998, Mart\'\i n et al. 2000), some have been
found in star forming regions like Taurus and Orion (e.g. Comer\'on,
Neuh\"auser \& Kaas 2000; Luhman et al. 2000), and a few have been
found as companions to main sequence or evolved stars (e.g. Nakajima
et al. 1996). Brown dwarfs are now also rapidly showing up in infrared
surveys of field stars like DENIS and 2MASS (e.g. Tinney, Delfosse \&
Forveille 1997; Kirkpatrick et al. 1999). The frequency of detection
in both cluster and field star surveys is now so high that brown
dwarfs may actually be as common as low mass stars. For a review, see
e.g. Tinney (1999) and Basri (2000).

With such high abundances, it becomes even more important to
understand the origin of brown dwarfs. It has been widely assumed that
brown dwarfs form a natural extension to normal stars, i.e. large
cores produce massive stars, and smaller cores produce lower-mass
stars (e.g. Elmegreen 1999). Alternatively, it has been suggested that
substellar objects may form through instabilities in circumstellar
disks (e.g. Pickett et al. 2000).  In a radical departure from these
views, we here suggest that brown dwarfs are stellar embryos which
have been ejected as a result of close dynamical interactions between
small unstable groups of nascent stellar seeds, i.e. brown dwarfs
differ from hydrogen burning stars only in that dynamical interactions
deprived them from gaining further mass by prematurely cutting them
off from their infalling gas reservoirs (Reipurth 2000). In this
paper, we examine this suggestion in some detail, and propose various
observational consequences and tests.

\section{DISINTEGRATING MULTIPLE SYSTEMS}

The dynamics of non-hierarchical triple or higher order systems has no
analytical solutions and the chaotic motions of the members can only
be followed numerically and analyzed statistically; for a review see
Valtonen \& Mikkola (1991). The motions can be divided into three
categories: {\em interplay}, in which the members move chaotically
among each other; {\em close triple approach}, when all members at the
same time occupy a small volume of space; and {\em ejection}, in which
one member departs the system after exchanging energy and momentum
with the two others. A close triple approach is a necessary but not
sufficient condition for ejection. Ejections often lead to escape, but
can also result in the formation of a hierarchical triple system, with
one body in an extended orbit. The remaining two members are bound
closer to each other, forming a tighter and highly eccentric binary
system. Most often, although not always, it is the lowest mass member
that is ejected; the escape probability scales roughly as the inverse
third power of the mass (e.g. Anosova 1986). The ejected member
acquires a velocity $ v_{eject} \sim 15~D_{ca}^{-1/2}$, where
$v_{eject}$ is in km~s$^{-1}$ and the closest approach $D_{ca}$ is in
AU (Armitage \& Clarke 1997).  Sterzik \& Durisen (1995, 1998) have
performed realistic numerical simulations of triple T~Tauri systems
and find ejection velocities of typically 3-4 km~s$^{-1}$ but with a
higher velocity tail. The decay of a triple system occurs
stochastically and can only be decribed in terms of the half-life of
the process. Anosova (1986) finds that within about one hundred
crossing times $t_{cr}$ almost all systems have decayed, where $t_{cr} \sim
0.17(R^3/M)^{1/2}$, and $R$ is a characteristic length scale for the
system in AU, and $M$ is the total system mass in $M_\odot$.

Three-body dynamical simulations including substellar members have
been performed by Sterzik \& Durisen (1999) and their dynamical
results can be summarized as {\em brown dwarfs are ejected because
they are of low mass}, in agreement with the general theory just
outlined.  This is in contrast to the evolutionary scenario advocated
here which can be summed up as {\em brown dwarfs are of low mass
because they are ejected}.

\section{COMPETITION BETWEEN ACCRETION AND EJECTION}
  
\subsection{The General Problem}

The mass of an individual star formed in a small multiple system is
largely determined by the competition of two independent processes: on
the one hand the accretion rate, which relates to the initial core
properties, and on the other hand the decay of the multiple system,
which leads to protostars being ejected and thus separated from
their natal mass reservoir.

  Consider a core containing a number of embryos whose initial mass is
much less than the mass of distributed gas in the core. The timescale
on which the bulk of the mass is accreted by the various embryos is of
order the dynamical timescale of the parent core (Bonnell et al.
2001a), i.e. a few times $10^5$ years for canonical core
properties. If the core mass divided by the number of embryos is
larger than the hydrogen burning mass limit, then evidently if
accretion went to completion {\it and} all embryos acquired their mass
equally, no brown dwarfs would result. Brown dwarf production
therefore requires at least one of the following conditions to be
realized: (i) some embryos are ejected before the accretion process
has gone to completion (i.e. the timescale for ejection is less than
the timescale on which the embryo would otherwise acquire a mass equal
to the hydrogen burning mass limit) and/or (ii) the embryos acquire
their mass very inequitably.

  A number of simulations, designed with rather different questions in
mind, can be used to shed light on each of these scenarios. 

In the first case, it is necessary for the ejection event to occur
rapidly (i.e. on a timescale of less than the $\sim 10^5$ year
timescale of the parent core). Therefore,  the relevant
embryos must form in a rather compact configuration (i.e. on a length scale
that is much smaller than the core size). For example, Sterzik \&
Durisen (1995) showed that the half-life for ejection is around $10^5$
years if three embryos occupy a volume of diameter around 200 AU (i.e.
on a size scale of around two orders of magnitude smaller than the
typical size of the parent core.) The system then consists of a
central mini-cluster of embryos which is fed mass from the infall of
the surrounding envelope. [To date, calculations of infall onto
multiple systems have been restricted to co-planar hierarchical
systems (Smith, Bonnell \& Bate 1997), and so are not relevant to the
current non-hierarchical situation].  Numerical simulations of
collapsing cores (e.g. Burkert \& Bodenheimer 1993) show that the
formation of a disk may be accompanied by the production of multiple
fragments in non-hierarchical orbits, whereas star-disk and disk-disk
interactions may likewise generate additional small fragments on the
size scale of disks (Boffin et al. 1998, Watkins et
al. 1998a,b). Fragmentation into even smaller volumes may be enhanced
by magnetic fields (Boss 2000).  Altogether, it would seem
straightforward to achieve the initial conditions for the ejection
model considered here.

   In the second case, it is not necessary that the embryos form in a
compact volume within the protostellar core, since an embryo does not
have to be ejected in order to avoid reaching the hydrogen burning
mass limit.  [One may instead envisage a situation similar to the
`prompt initial fragmentation' model proposed by Pringle (1989) in
which fragments are seeded by existing structure in the core].  In
this case, it is enough if more gentle encounters perturb an embryo
into an orbit where it does not intersect the densest part of the mass
reservoir. Such a situation has an inbuilt feedback, since once an
embryo has fallen behind in the race to acquire mass, it is more
readily deflected away from the central densest regions. Likewise,
embryos which acquire an early start in the mass race remain in the
centre and preferentially intersect the infalling mass flow. This
runaway to disparate masses is the basis of the competitive accretion
scenario explored in a series of papers by Bonnell and collaborators
(Bonnell et al. 1997, 2001a, 2001b). Of greatest relevance to the
present discussion of small group dissolution are the simulations
contained in Bonnell et al. (1997) which follow the evolution of cores
containing 5-10 embryos initially distributed throughout the core
volume. The trajectories and mass acquisition histories of the embryos
clearly demonstrate the evolutionary path described above. Although
these are only pilot simulations, and cannot be used to draw
statistical inferences about the resulting mass distributions of
stars, it is notable that the simulations do show a significant number
of stars acquiring less than $10 \%$ of the mean mass of stars in the
clusters (corresponding roughly to the mass range of brown dwarfs:
see, for example, Figure 2 of Bonnell et al. 1997 in which 2 out of
the 10 objects end up in this category).

  Although in the second case, an embryo can remain low mass without
being ejected (on the mass acquisition timescale of $\sim 10^5$
years), the non-hierarchical situation is ultimately unstable, so that
on timescales not much longer than this the system will break up into
a system of binary and single stars. Thus in either case, the expected
outcome is the production of a population of `stars' whose mass is
much less than the average mass produced by the core - i.e. objects we
might loosely identify with brown dwarfs and very low mass stars. The
critical difference between these two cases is the expected ejection
velocity of the "brown dwarfs": for the case of the initially compact
embryo configuration, the typical separations of close encounters are
two orders of magnitude less than for the case that embryos are
initially distributed throughout the core, and this translates into
ejection velocities that are an order of magnitude greater (see
Sects. 4.1 and 4.4).

 The rough estimates presented here suggest that brown dwarfs could be
common, maybe as common as normal stars, which is indeed what some
recent surveys suggest.  However, the only proper way to estimate the
efficiency of this mechanism for producing brown dwarfs is through
detailed numerical simulations of the dynamical evolution of small
N-body systems in a time-variable potential and including the presence
of circumstellar material. Though a number of simulations have
attempted to capture aspects of this complex problem, there is clearly
much more to be done in this area.

\subsection{A Toy Model}

To get a feeling for the time scales involved, consider the following
order-of-magnitude estimates. Assume a flattened cloud is collapsing
with a time-averaged mass infall rate $\dot{M}_{infall} \sim 6 \times
10^{-6}(T/10 K)^{3/2}$ M$_\odot$ yr$^{-1}$, as derived by Hartmann,
Calvet \& Boss (1996). A 1~M$_\odot$ cloud at 10~K would then take 1.7
$\times$ 10$^5$ yr to collapse. Assume that the collapse fragments
into a number $N_{mul}$ of stellar embryos and that the infalling mass
is distributed among them according to their mass ratios $q_i =
M_i/M_{prim}$ (which may be time-dependent), where $1 \leq i \leq
N_{mul}$, and $M_{prim}$ is the time-dependent mass of the most
massive embryo.  Further assume that a part $f$ of the infalling gas
is lost in outflow activity, where we adopt $f$$\sim$0.3. The growth
rate of the $i$th embryo is then $\dot{M}_{i} = \dot{M}_{infall}
\times (1-f) \times q_i/N_{mul}$, and for the simplest case where
$N_{mul} = 3$ and $q_i = 1$ it will then take a typical embryo about 6
$\times$ 10$^4$ yr to grow beyond a substellar mass.

If the three stellar embryos occupy a volume with diameter
200~AU, and if we consider a total mass in the range 0.06 to 0.24
M$_\odot$, then the characteristic crossing time $t_{cr} \sim
0.17(R^3/M)^{1/2}$ is between 1000~yr and 2000~yr.  If, on the other
hand, the configuration is tighter, occupying a volume with a diameter
of only 20~AU, then $T_{cr}$ is merely 30-60~yr.

The usual decay equation is $ n_t/n_o = exp(-0.693t/\tau) $, where
$\tau$ is the half-life of the decay.  In their numerical simulations,
Sterzik \& Durisen (1995) found that about 95\% of all systems had
decayed after about 100 crossing times. This allows us to link the
crossing time and the half life of the decay, $ \tau = 23.1t_{cr}$, so
that to first order $ n_t/n_o = exp(-0.18 \dot{M}_{infall}^{1/2}
(1-f)^{1/2} N_{mul}^{-1/2} R^{-3/2} t^{3/2})$.

While the stellar embryos are still very small, the crossing time, and
therefore the half life, is extremely long, i.e. the embryos do not
effectively start to interact until a time T$_i$ when they have a
certain mass. When the embryos are very small, their awareness of each
other is limited because they are surrounded by the massive infalling
envelope, and this further adds to $T_i$.  But as the envelope
material thins out and the embryo masses increase, their dynamical
interactions become important. A more precise value of $T_i$ can only
be determined by numerical experiments.

We will say that there is a reasonable chance that a brown dwarf is
ejected if the half-life $\tau$ of the decay is less than
$T_{*}~-~T_i$, where $T_{*}$ is the free-fall time required to
build up a multiple system of objects of average mass 0.08 M$_\odot$,
i.e. $T_{*} = (N_{mul} \times 0.08)/\dot{M}_{infall}$. If the ejection
occurs at a time later than $T_{*}$, then the ejected object is a
star. In either case it has been assumed that the amount of
circumstellar material the ejected object brings with it is small.

Figure~1 shows a simple, schematic presentation of these processes in
a growth vs decay diagram, based on the values already discussed. The
stapled line represents the growth of an embryo, which reaches the
stellar mass threshold after $6 \times 10^4$ yr. The solid curve shows
$n_t/n_o$, which is a measure of the probability that the multiple
system has not yet decayed.  To be conservative, we have assumed the
embryos occupy a volume of diameter 200~AU, resulting in a large
$t_{cr}$.  The time when interactions begin to take place, $T_i$, has
arbitrarily been set to $3 \times 10^4$ yr. In the particular example
shown, about one third of a population of triple systems would have
decayed before the expulsed member reached the stellar mass limit.

An important issue concerns the rate at which mass is accreted.  The mass
accretion rate adopted in Figure~1 is constant, in accordance with Shu
(1977), but studies of shock-induced collapse suggest that infall
might start with a powerful burst, briefly reaching values as high as
10$^{-4}$ M$_\odot yr^{-1}$, and subsequently decreasing (e.g. Boss
1995). If so, an embryo might much more quickly reach and exceed the
stellar mass limit, leaving less time for the ejection of brown
dwarfs. 

This is, however, {\em not} a problem for the ejection scenario,
because each of the following four physical processes will allow more
time for ejection. Firstly, the infalling mass is distributed among
the stellar seeds, and if $N_{mul}$ is higher than three, $T_{*}$
becomes correspondingly higher. Secondly, in the calculations above we
have for simplicity assumed that the components of newborn multiple
systems have equal access to growth from infalling material, but as
discussed in Sect.~3.1, competitive accretion is likely to
be a very important process (Bonnell et al. 1997), leading to $q_i$
different from 1, and this would leave one or more embryos as
substellar objects for extended periods of time. Third, if embryos are
born in a more compact rather than a wide configuration (Sect.~3.1), then
the crossing time decreases dramatically, leading to enhanced
probability of very rapid ejection, on timescales of just a few thousand
years. Finally, it is conceivable that stellar embryos may be formed
over a period of time, with late-forming embryos being rapidly ejected
by their more evolved siblings.

\section{OBSERVATIONAL CONSEQUENCES OF THE EJECTION MODEL}

In the following, we outline a number of issues relevant to current
studies of brown dwarfs as seen from the perspective of the ejection
model.

\subsection{Detecting Brown Dwarfs near Embedded Sources}

All stars, as they form initially as stellar embryos and rapidly gain
mass through infall, must at some early point pass from substellar to
stellar mass objects. Therefore, if isolated brown dwarfs are the
result of the disintegration of a system of multiple stellar embryos,
the moment of decay must take place very early in the infall process,
and it thus appears likely that brown dwarfs are ejected during the
Class~0 phase.

It follows that {\em in order to find the very youngest brown dwarfs
one should search near Class~0 sources}.  The number of Class~0
sources known is limited, and suggests that any attempt to compare
decay calculations with observations will be severely limited by small
number statistics. To get a practical sense of the observed separation
between a nascent brown dwarf and its siblings, assume that it is observed
at a time $t$ [yr] after ejection and moving with the space velocity
$v$ [km s$^{-1}$] at an angle $\alpha$ to the line-of-sight in a star
forming region at a distance $d$ [pc]. Then the projected separation
$s$ in arcsec is $ s = 0.21~v~t~d^{-1}~sin\alpha. $

As an example, assume that a brown dwarf is moving out of a nearby
($d$ $\sim$ 130 pc) cloud at an angle of 60$^o$ to the line-of-sight
with a space velocity of 3 km s$^{-1}$. At a time of only 10$^4$ yr after
ejection, the brown dwarf is already 42 arcsec from its site of birth,
and after 10$^5$ yr, when the embedded phase of the original Class~0
source is about to end, the brown dwarf is 7 arcmin away, and its
origin is soon lost in the mist of time.

Half of all ejected brown dwarfs will move into the cloud from which
they formed. Such objects will be detectable only as highly extincted
and weak infrared sources.

  The rather high ejection velocities quoted above correspond to brown
dwarfs ejected from a compact configuration (see Sterzik \& Durisen
1998) and one sees that one has little chance of associating such
ejected objects with the Class 0 star at an age comparable with the
average age of Class 0 sources. If, however, brown dwarfs are ejected by
the softer interactions envisaged in the competitive accretion
scenario (see Section 3) then their ejection velocities are typically
smaller, due both to the dissipative effects of gas drag and to the
larger mean encounter distances in this case.  The ejection velocities
have not been quantified in this case, but rough estimates suggest
that they may be only of order a few times $0.1$ km~s$^{-1}$.  In this
case, therefore, one would expect to detect one or more brown dwarfs
around any given Class 0 object.

\subsection{Binary Brown Dwarfs}

As more brown dwarfs become known and more detailed studies of
individual objects are undertaken, the number of known brown dwarfs
paired in binaries is certain to increase (e.g. Mart\'\i n et
al. 2000; Reid et al. 2001). Among the first identified bound pairs of
brown dwarfs is PPl~15, which has an orbital period of six days (Basri
\& Mart\'\i n 1999), and other pairs have been recently reported
(e.g. Mart\'\i n, Brandner \& Basri 1999).
 
 Intriguingly, there is gathering evidence that whereas close pairs
may be common (comparable with the binary fraction for solar type
stars), there is a striking absence of wider pairs: to date, no brown
dwarf binaries have been detected with separations larger than 23~AU,
despite a lack of any obvious selection bias against the detection of
such binaries (Mart\'\i n et al. 2000).  The fact that this apparent
dearth of wider brown dwarf pairs is found both in the field and in
clusters, hints that its origin is not due to the larger scale
environment, but may reflect smaller scale processes that are common
to both environments. In the context of the present paper, we would
argue that the small scale clustering is ubiquitous (regardless of
whether the parent regions survive as large scale bound clusters) and
that any observational signatures of this origin would be expected in
both field and cluster environments.  [Note that although some young
clusters such as the Orion Nebula Cluster show no hint of primordial
sub-clustering (Bate, Clarke \& McCaughrean 1999), there is ample time
for such clustering to have dissolved over the ~ 2 Myr age of the
cluster (Scally, in preparation)]

    It is clear, however, that purely point mass dynamical
interactions will not produce brown dwarf binaries. Such interactions
give rise to a dynamical biasing (McDonald \& Clarke 1993), such that
the resultant binary is usually composed of the most massive two
members of the mini-cluster (van Albada 1968). McDonald and Clarke
showed that under this assumption binaries comprising two brown dwarfs
would essentially never be formed, a result confirmed by the more
recent numerical simulations of Sterzik \& Durisen (1998).

   The situation is, however, different in the case that cluster
dissolution is associated with dissipative interactions between the
stars (embryos).  An obvious source of such dissipation is provided by
the presence of extended disks around each object, which exert a drag
on neighboring stars whose orbits intersect the disk (Clarke \&
Pringle 1991a, 1991b, Hall, Clarke \& Pringle 1995).  McDonald \&
Clarke (1995) modeled the disintegration of small multiple systems in
which the dynamics was crudely modified by parameterized star-disk
interactions.  They showed that the dynamical bias is considerably
weakened in this case: dissipative interactions harden the resulting
binaries and allow for the survival of low mass pairs that would
otherwise have been broken up in dissipationless
environments. Nevertheless, the predicted binary fraction for brown
dwarfs in these simulations is still low (around $5 \%$).  More
realistic calculations (Delgado, in preparation) are required in order
to investigate whether this represents a serious objection to the
ejection model.

  Whatever the origin of brown dwarf binaries in mini-clusters, they
clearly have to be rather close in order to survive the ejection.
More specifically, Sterzik \& Durisen (1998) found that the binaries
that survive had separations not exceeding roughly one third of the
mini-cluster radius. The apparent observed dearth of binary brown
dwarfs with separation greater than $20-30$~AU would then certainly be
consistent with the scenario that brown dwarfs were ejected from a
compact configuration (i.e. within a volume of radius $\sim 100$ AU).

  A further criterion necessary for the survival of brown dwarf
binaries is that after formation they do not continue to accrete
beyond the hydrogen burning mass limit. In order for this to be the
case, it is probably necessary for the interaction giving rise to the
binary to eject it in the process. Again, further simulations will
clarify if this is feasible.

\subsection{Brown Dwarfs as Companions to Normal Stars}

Many of the earliest searches for brown dwarfs were done towards main
sequence stars or white dwarfs, and their mostly negative outcome
added to the early speculations that brown dwarfs are rare. We now
know, of course, that brown dwarfs are not rare, only as companions to
other stars, the ``brown dwarf desert'' (Marcy \& Butler 1998).
Complete references to these searches are given by Basri (2000), who
notes that the incidence of brown dwarf companions to stars with
masses of 0.5 M$_\odot$ or more does not exceed 1\%.

If brown dwarfs are formed as stellar embryos which were prematurely
ejected, this observational result is readily explained. If these
stellar embryos had {\em not} been ejected, they would have continued
to accrete matter from the infalling gas, and eventually would have passed
from the substellar to the stellar regime. Indeed, in the ejection
model, the problem is not to explain the absence of brown dwarfs
around more evolved stars, but rather to explain how any brown dwarf
could be still bound in such a system, because continued accretion on
to both of the binary components would raise the brown dwarf companion
above the hydrogen burning mass limit. There are two different
approaches to this.

In the first scenario, we must assume that binaries consisting of a
brown dwarf and a main sequence star must have been ejected very early
from the mini-cluster, in precisely the same manner discussed in
Sect. 4.2 for pairs of brown dwarfs, except that in this case one of
the components happens to have already exceeded the hydrogen burning
limit by the time of ejection. Since statistically the lightest
members of a mini-cluster are more likely to be ejected, it follows
that binaries with a main sequence star and a brown dwarf are more
likely to have a very low mass star like an M dwarf as a member than a
more massive star like a G dwarf. Within the limited statistics
available so far, this indeed seems to be the case (Rebolo et
al. 1998).

In the second scenario, competitive accretion in a mini-cluster
creates a group of objects with a wide range in masses. As the small
N-body system dissolves through dynamical interactions, the formation
and subsequent ejection of binary systems may be influenced by a
dynamical bias against binaries with very low mass companions.  In the
case of purely point mass gravitational encounters, brown dwarfs
should rarely be companions to normal stars (McDonald \& Clarke 1993),
since the relatively fragile nature of such extreme mass ratio
binaries makes them vulnerable to exchange reactions with other more
massive stars in the newborn multiple system. The observation of at
least a small number of such pairs might imply a role for
dissipative interactions which harden the resulting binary and permit
its survival: in the simulations of McDonald \& Clarke (1995), which
employed very massive and extended disks, large numbers of low mass
stars or brown dwarfs ended up as companions to solar type stars. The
fact that, observationally, such pairs are {\it not} plentiful
suggests a more limited role for dissipative interactions in real
systems.

Because the lowest mass members of a multiple system have a much
increased chance of being ejected, it follows that if giant planets
orbiting other stars were formed as stellar embryos, then they should
be exceedingly rare. This is obviously contrary to the observation of
numerous giant planets around stars, and it suggests that brown dwarfs
and giant planets must form in two different processes.

\subsection{Kinematics of Brown Dwarfs}

   If brown dwarfs are ejected stellar embryos, they should carry
kinematic evidence reflecting their origins in small multiple
systems. To first order, the ejected member from such a system
acquires a velocity comparable to the velocity attained at pericenter
in the close triple encounter. Low mass ejected members will therefore
generally have higher velocities than higher mass objects.

 Whether such an ejection history is likely to be detectable, through
radial velocity and proper motion measurements of brown dwarfs,
depends on the magnitude of the ejection velocities (which themselves
depend on the closeness of triple encounters) and the general level of
stellar motions in the star forming complex. For example, in clusters
such as Orion or the Pleiades, where the velocity dispersion of stars
in the large scale cluster potential is several km s$^{-1}$, brown
dwarfs ejected at such speeds from primordial mini-clusters would
leave no kinematic imprint. Note that the very existence of brown
dwarfs in the Pleiades means that retained brown dwarfs must have
been ejected at less than the cluster's escape velocity; in practice,
even in the compact simulations of Sterzik \& Durisen (1998), most
brown dwarfs would be retained in the Pleiades. There may, however, be a
significant tail of brown dwarfs ejected at higher velocities, which
would populate the outer reaches of the Pleiades. Numerical
simulations of the Pleiades that do {\it not} include primordial
sub-clustering (Fuente Marcos \& Fuente Marcos 2000) indicate that
very few brown dwarfs are expected to be ejected by subsequent
encounters within the Pleiades. Hence the detection of a brown dwarf
halo around the Pleiades (or objects with discrepant proper motions)
would provide unambiguous evidence of their origin in very compact
mini-clusters.  Further brown dwarf searches in the vicinity of the
Pleiades (e.g. Bouvier et al., in preparation) are therefore likely to
be particularly valuable.

 Further constraints on the kinematic history of brown dwarfs are
provided by their spatial distributions within star formation
regions. For example, Hillenbrand \& Carpenter (2000) have shown that
in the core of the Orion Nebula Cluster, the mass function turns over
at around 0.15 M$_\odot$. Given that there is evidence for mass
segregation amongst more massive stars in the ONC (Hillenbrand \&
Hartmann 1998), it is obviously of great interest to determine if the
substellar mass function has a similar form at larger radii in the
cluster.  In the case of Taurus (where the stars are not mutually
bound) it is tempting to speculate that the paucity of brown dwarfs
recently reported by Luhman et al. (2000) may be due to their higher
ejection velocities, so that they have already left the region within
which the bulk of T Tauri stars are currently concentrated.

It should be emphasized that brown dwarfs in very massive and/or very
old clusters have velocities that are higher because their kinematics
have been dominated by two-body relaxation. Therefore, such clusters
should {\em also} be surrounded by a halo of brown dwarfs, but not
because of their high ejection velocities. A proper test of the ejection
scenario is made only by finding a halo of brown dwarfs around
clusters small enough or young enough that relaxation has not yet
dominated their kinematics.  With regard to old brown dwarfs in the
field, these have not been dynamically relaxed, but the majority must
have evaporated from large clusters, in the process losing their
original kinematic history (Spitzer \& Harm 1958).

\subsection{Substellar Equivalents to the Classical T Tauri Stars}

 Classical T Tauri stars show emission line characteristics and
veiling which identify them as young accreting objects (Herbig
1962). The source of this accretion is a circumstellar disk. If brown
dwarfs are ejected stellar embryos, then their disks are likely to be
pruned during the close interactions that ejected them. Armitage \&
Clarke (1997) showed that very close encounters, which truncate the
disks to radii of a few AU, could promote the rapid decline of
classical T Tauri star characteristics thereafter and one might on
these grounds expect that all but the very youngest brown dwarfs are
unlikely to show any of the spectroscopic signatures of infall and
outflow which we have come to associate with extremely young stellar
objects. This appears to be consistent with current observations of
young brown dwarfs (Mart\'\i n 2001). In the case of ejected brown
dwarfs, an empirical upper limit on expected encounter separations may
be provided by the observed lack of brown dwarf binaries at
separations greater than $20-30$ AU. Further modeling is required in
order to quantify how quickly accretion characteristics would be
expected to decline following brown dwarf ejection: we, however, point
out that because of the draining of the pruned disk, and because
ejected stars have no source of disk replenishment, it may well be
difficult to find the substellar equivalents of classical T Tauri
stars, except perhaps as newly ejected objects located in the vicinity
of Class 0 sources.

 A further issue is the effect of ejection and disk pruning on the
rotation of brown dwarfs, since in many models the disk plays the
major role in braking young stars (K\"onigl 1991, Cameron \& Campbell
1993, Armitage \& Clarke 1996).  One might thus expect that ejected
brown dwarfs in star forming regions should be rapidly rotating
(i.e. at speeds close to break up). However, it turns out that
extremely close encounters, even at the sub-AU level, are required in
order to significantly affect disk braking, at least in the case of
more massive stars (see the discussion in Clarke \& Bouvier
2000). Further calculations using brown dwarf parameters and plausible
enounter distances, are required before one can make firm predictions
of the importance of this effect, although the {\it sign} of the
prediction (a tendency towards more rapid rotation) is clear.

\subsection{The Low-Mass End of the IMF}

Numerous studies have explored the origin of the initial mass function
(see e.g. Kroupa 1995; Larson 1999; Elmegreen 2001). Such analyses
have been greatly helped by the increasingly accurate determinations
of the mass function reaching into the very low mass and substellar
regimes (e.g. Reid et al. 1999, Luhman et al. 2000, Najita et
al. 2000, Hillenbrand \& Carpenter 2000, Muench, Lada \& Lada
2000). These studies have firmly established that the standard
Salpeter power law slope of the IMF at higher masses gives way to a
much shallower slope at lower masses or even a turn-over.  While the
main power law slope probably is related to the hierarchical
structure of interstellar clouds, the flat part of the IMF at lower
masses may result from a second, unrelated, physical process
(e.g. Elmegreen 2000).

The ejection model can naturally produce the observed turn in the IMF .
With reference to Figure~1, the shape of the very low mass end of the
IMF is significantly influenced by the difference between $T_i$ and
$T_{*}$. If $T_i$ should be greater than $T_{*}$ all objects ejected
would be low mass stars. If interactions occur even just slightly
before $T_{*}$, the steep decline of $n_t/n_o$ ensures that a
population of brown dwarfs will come into existence. And if $T_i <<
T_{*}$, brown dwarfs will be abundant.

\section{DO LOW MASS STARS FORM IN SMALL MULTIPLE SYSTEMS?}

In recent years it has been increasingly well documented that binarity
is common among pre-main sequence stars and probably higher than among
main sequence stars (e.g. Reipurth \& Zinnecker 1993, Ghez,
Neugebauer \& Matthews 1993, K\"ohler \& Leinert 1998). More recently,
an analysis of deeply embedded outflow sources has revealed an {\em
observed} binary frequency of around 80\%, with half in higher order
systems (Reipurth 2000).

On the theoretical side, there is growing consensus that fragmentation
during the protostellar collapse phase is the principal mechanism for
forming binaries and multiple systems (e.g. Bodenheimer et al. 2000,
Bonnell 2001, Whitworth 2001), although the formation of close binary
stars may require additional steps, as for example dynamical
interactions between the nascent stars.

For brown dwarfs to be formed as ejected stellar embryos, close triple
approaches must occur very early in the infall process. This is
greatly facilitated if from the outset stellar seeds are formed close
to each other. In this context it is of interest that the latest
three-dimensional calculations of fragmentation in a collapsing core
including magnetic fields show that magnetic tension helps avoiding a
central density singularity (Boss 2000). These models suggest that
fragmentation leads to a transient quadruple system. This is the
situation which is ideal for dynamical interactions resulting in the
early expulsion of a stellar embryo.

Dynamical interactions between members of small multiple systems are
well studied and it is incontrovertible that they sometimes must play
a role during the star formation process. The issue is therefore not
{\em whether} stellar embryos can be ejected from a nascent multiple
system, but rather {\em how often} this process occurs. Larson (1972,
1995) has suggested that all stars may be born in binary or multiple
systems, and the high observed multiplicity frequency of Class~0/I
sources near $\sim$80\% would seem to support such a scenario
(Reipurth 2000). The detection and statistical study of newborn brown
dwarfs in star forming regions may therefore cast light on the general
question of how low mass stars form.

\section{CONCLUSIONS}

 We have argued here that the primary route for brown dwarf formation is
{\it not} through the collapse of low mass cores, but that rather they
are ejected as a result of encounters in multiple systems comprising 
a small number of stellar embryos. A number of simulations give plausibility
to this picture, though the expected ejection velocities (and other properties
such as the presence of accretion diagnostics, the range of binary separations
and brown dwarf rotation rates) will depend on whether the embryos form on
size scales comparable with the parent core ($ \sim 10^4$ AU; `prompt initial
fragmentation') or on size scales of protostellar disks ($\sim 100$ AU). 

  To date, simulations have tended to address the problem in piecemeal
fashion, focusing separately on the roles of point mass dynamics, star-disk
interactions and competitive accretion. A clear goal for theorists is more
realistic simulations that integrate all these features, a goal that is
becoming increasingly realizable with the aid of advances in massively
parallel computer architecture. In the absence of such simulations,
the predictions given above have necessarily been laced with numerous
caveats. Rather than repeating these here, we set out the key observations
in the coming years that will provide the datasets necessary to test the
models:

 1. {\em Brown dwarf searches in the vicinity of Class 0 sources}

 A null result in this area would not contradict the ejection model,
but would imply high ejection velocities (several km s$^{-1}$)
consistent with embryos being ejected from a compact configuration
($\sim 100$ AU). A positive result would provide strong evidence in
favour of the hypothesis, but, if common, would imply rather slow
ejection velocities.

 2. {\em Brown dwarf binaries}

  The apparent upper limit of $20-30$ AU on the separation of brown
dwarf binaries may contain an important fossil record of a history
of close encounters. Hence the placing of this result on a firmer statistical
basis is an important goal. Likewise, it is important to establish whether
close brown dwarf pairs are indeed as common as current estimates suggest.
Although the survival of such pairs is unproblematical in the
ejection model, it is not clear whether it is easy to form such low
mass pairs starting from non-hierarchical initial conditions.

 3. {\em Brown dwarfs as companions to normal stars}

 The result that such companions are rare, at least in the case of solar
type stars, is now well established. An important goal for simulations
will be to discover whether they can simultaneously reproduce this
property together with a high fraction of  brown dwarfs as binary
primaries (point 2. above).

  4. {\em Brown dwarf kinematics and distributions around open
clusters}

  The presence of brown dwarfs in open clusters places an upper limit
of several km s$^{-1}$ on the ejection velocities of brown dwarfs that
are retained in the cluster. Brown dwarf searches in the vicinity of
the cluster will, however, explore a possible tail of higher velocity
ejections. The existence of such a brown dwarf halo around the
Pleiades, for example, would provide strong evidence in favour of the
ejection hypothesis, since simulations starting from smooth initial
conditions (no mini-clusters) do not eject brown dwarfs to large
distances. On the other hand, a null result would not rule out the
ejection hypothesis but would constrain the velocity distribution of
ejected objects.

 5. {\em Circumstellar diagnostics and rotation rates of young brown dwarfs}

  Ejected brown dwarfs have suffered encounters that will have pruned
their circumstellar disks. In consequence, one might expect weaker
accretion diagnostics and faster rotation in young brown dwarfs
compared with young stars. Further work is required to quantify the
predicted magnitude of these effects, whilst considerably more
observational data is needed in this area.

\vspace{0.5cm}

\acknowledgments

We thank John Bally, Jerome Bouvier, Nick Gnedin, Richard Larson,
Eduardo Mart\'\i n, and Michael Sterzik for discussions and helpful
comments, Ka Chun Yu for preparing the figure, and Alan Boss for a
helpful referee report.

\clearpage



\begin{figure*}[tbp]
\includegraphics[height=3.5in]{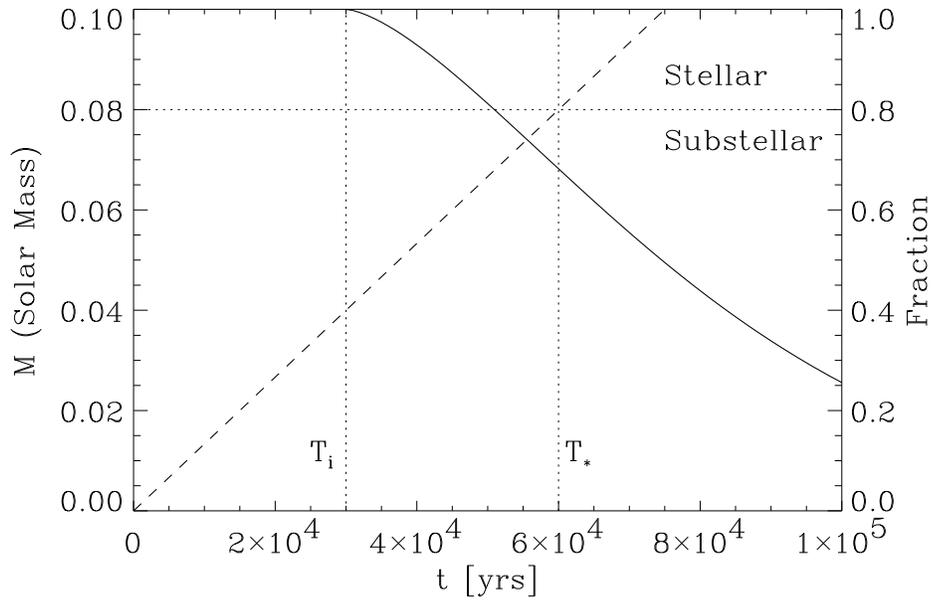}
\caption{ This growth vs decay diagram provides a schematic
presentation of two competing processes, accretion of mass and the
probability of dynamical ejection, affecting a triple system of
forming stellar embryos. The stapled line indicates the growth of an
embryo (assuming three embryos share a constant infall rate of
$\dot{M} = 6 \times 10^{-6} M_\odot$ with a part $f\sim$0.3 being lost
to outflow), and the solid curve indicates $n_t/n_o$, the fraction of
still intact systems, or equivalently the probability that a given
system has still not decayed. $T_{*}$ is the time when the embryo has
reached a mass of 0.08~$M_\odot$, while $T_i$ is the (poorly
constrained) time, when the embryos have enough mass to begin
dynamical interactions.}
\end{figure*}

\end{document}